\documentclass[runningheads]{llncs}
\usepackage[utf8]{inputenc}
\usepackage{amsmath}
\usepackage{amsfonts}
\usepackage{amssymb}
\usepackage{graphicx}
\usepackage{hyperref}
\usepackage[numbers,sort&compress]{natbib}

\usepackage[T1]{fontenc}
\usepackage{graphicx}
\usepackage{float}

\usepackage{booktabs}
\usepackage{caption}
\usepackage{array}
\usepackage{comment}
\usepackage{xcolor}
\usepackage[font=small, skip=10pt]{caption}

\usepackage[a4paper, left=37.5mm, right=37.5mm, top=37.5mm, bottom=37.5mm]{geometry}

\usepackage[backgroundcolor=white,textsize=tiny]{todonotes}
\DeclareUnicodeCharacter{0301}{\'{e}}
\usepackage{multirow}

\begin{document}
\title{Shifting Patterns of Extremist Discourse on Facebook: Analyzing Trends and Developments During the Israel-Hamas Conflict}

\titlerunning{Analyzing Trends and Developments During the Israel-Hamas Conflict}

%
%

\author{Rr. Nefriana\inst{1} \and
Muheng Yan\inst{1} \and Ahmad Diab\inst{1} \and Wanhao Yu\inst{2} \and
Deborah L. Wheeler\inst{3} \and Andrew Miller\inst{3} \and Rebecca Hwa\inst{4} \and Yu-Ru Lin\inst{1**}}
\authorrunning{Nefriana et al.}
%
\institute{ University of Pittsburgh, Pittsburgh, PA 15260, USA\\
\email{\{rr.nefriana, muheng.yan, ahd23, yurulin\}@pitt.edu}
\footnote{Corresponding author.}
\and
University of North Carolina at Charlotte, Charlotte, NC 28223, USA\\
\email{wyu6@charlotte.edu}
\and
United States Naval Academy, Annapolis, MD 21402, USA\\
\email{\{dwheeler, millera\}@usna.edu}
\and
George Washington University, Washington, DC 20052, USA\\
\email{rebecca.hwa@email.gwu.edu}
}
\maketitle              
\begin{abstract}
This short paper explores trends in extremist Facebook data from July 2023 to June 2024. We examined engagement, sentiment, and topics within Facebook groups categorized as anti-Israel/Semitic, anti-Palestine/Muslim, and anti-both, mapping these trends against five major events related to the recent Israel-Hamas conflict. Our findings support the hypothesis that shifts in trends correspond with these key events, showing varying patterns across different group categories. We observed decreased activity proportion in anti-both groups and increased activity proportion in the two one-sided hate groups at the conflict's onset. This pattern reversed after the Israeli troop withdrawal from Khan Yunis, Gaza. During the conflict, negative content proportion surged, and neutral content proportion fell in all the three group categories. Anti-Palestine/Muslim groups' discourses shifted from religious to social media activism and political/protest around the time the war began, while anti-Israel/Semitic groups moved from political/protest to religious topics a couple of weeks before the war.

\end{abstract}
\def \rev #1{{\color{black}{#1}}}

\newcommand{\todoo}[1]{{\small\textcolor{blue}{[ToDo: #1]}}}

\section{Introduction}

The Israel-Palestine conflict often sparks intense global debates among ethnic groups, political parties, and sovereign states \cite{gelvin2014israel}. Social media has extended these discussions beyond geographical limits, enabling global engagement and activism. However, this expansion also amplifies hate speech, particularly against Muslim and Jewish communities \cite{castano2021internet}. Anti-Muslim and anti-Semitic groups frequently exploit major events \cite{kramer2006anti}, with the Hamas-led war in October 2023 fueling increased hatred. This highlights the need for a detailed examination of social media interactions related to anti-Muslim and anti-Semitic content during significant geopolitical events.

Hate speech targeting Muslims and Jews has distinct historical and political roots \cite{berecz2017relevance}, but the Israel-Palestine conflict intertwines them. Anti-Muslim narratives often exploit Islamophobic sentiments, depicting Muslims as threats to Western values and security \cite{berecz2017relevance}. Conversely, anti-Semitic hate usually draws on long-standing conspiracy theories about Jewish influence in politics, economy, and culture \cite{berecz2017relevance}. The attacks initiated by Hamas and the ensuing political-military aftermath by Israel appear to align with and intensify these existing prejudices, further polarizing online discourse.

Understanding hate speech and social media engagement during the Israel-Palestine conflict is crucial for several reasons. It helps identify patterns in the rise and fall of online hate speech, offering insights into how major events can either provoke or reduce hatred. This knowledge can inform the development of more effective strategies to counteract online hate by pinpointing key moments for intervention. Additionally, it aids in anticipating potential spikes in hate speech, allowing for timely and proactive measures to mitigate harm.

Despite extensive research on online hate speech \cite{castano2021internet}, there are still gaps in understanding the specifics of hatred during the 2023 Israel-Palestine war. First, the variation in hate speech engagement throughout the conflict needs more investigation. It is important to determine if spikes in engagement align with specific events—military, political, or diplomatic—and to analyze the patterns of these fluctuations. Second, the link between major events in the conflict and changes in online sentiments within anti-Muslim and anti-Jewish groups is underexplored. While previous studies indicate that sentiments can shift rapidly in response to news and events \cite{chetty2019trigger}, the direction and correlation of these changes with specific incidents have not been studied in the context of mutual hatred in a conflict. Lastly, examining how discussion topics on social media evolve during the conflict can provide additional insights. Identifying shifts in focus—such as reports on attacks, religious debates, empathy for war victims, or condemnation of perpetrators—can help understand how hate groups sustain and expand their influence.

To address this gaps, this study proposes the following research questions:
\begin{itemize}
    \item \textbf{RQ1}:How does engagement with hate speech vary as the Israel-Palestine conflict develops? Are these changes related to major events?
    \item \textbf{RQ2}: How do sentiments within anti-Muslim and anti-Jew groups change during the conflict? Are these changes related to major events?
    \item \textbf{RQ3}: How do the topics of social media discussions change within these groups over time?
\end{itemize}

This research provides an understanding of the dynamic of online hate speech during the Israel-Palestine conflict, which will contribute to the broader discussion on online hatred to promote a more inclusive digital space.

\section{Related Works}

Hate speech on social media is a pervasive issue, amplified by the reach and anonymity these platforms offer. Online hate speech, which disparages individuals based on characteristics like race, religion, or gender, is facilitated by the unregulated nature of social media \cite{zhang2019hate}. The prevalence is alarmingly high, with 80\% of people in the EU encountering it and 40\% feeling personally attacked or threatened \cite{gagliardone2015countering}. This contributes to an environment of prejudice and intolerance, which can escalate to violence \cite{papacharissi2004democracy}. Platforms like Facebook and Twitter play a key role in mediating and amplifying hate speech \cite{castano2021internet}. Their design and policies affect the frequency and spread of hate speech, with anonymity features protecting harassers and algorithmic suggestions promoting racist content \cite{ben2016hate}.

Hate speech online can manifest as religious hate speech, racism, political hate, and gendered hate \cite{castano2021internet}. For example, anti-Islam hate, a focus of this study, is prevalent on social media and often associated with Islamophobia and global cultural processes \cite{horsti2017digital}. Violent events, such as terrorist attacks, are linked to surges in online hate speech; for instance, the \#StopIslam hashtag trended on Twitter after the Brussels attacks in 2016, spreading racialized hate speech against Muslims \cite{poole2019contesting}. Geo-political events also impact online hate speech, with the Israeli-Palestinian conflict being a key example. During intensified conflicts, like military operations in Gaza, there is a noticeable increase in anti-Semitic content. This pattern is consistent in various European countries, where anti-Semitism and anti-Zionism often overlap, complicating the distinction between political criticism and outright hate \cite{berecz2017relevance}. The escalation of hostilities in the Israel-Hamas conflict contributes to the spread of hate speech, with online platforms filled with content reinforcing existing prejudices and stereotypes \cite{berecz2017relevance}.

Given these dynamics from past research, we build an understanding of how such conflict-driven hate speech would evolve. This study aims to fill the gaps in existing research by examining the trends in engagement, sentiment, and topics within extremist Facebook groups throughout the Israel-Hamas conflict.
\section{Methodology}

\subsection{Data}

We examined Facebook data to answer the research questions. To create a corpus of hate speech against Jews and Muslims, we start with a list of keywords curated by an expert on Middle Eastern regional politics.
This list contains phrases and slogans that explicitly express hate against either Muslim or Jewish groups. This list includes explicit phrases like "death to Israel" and "Arabs are inferior," as well as conflict-related terms such as "From the Rivers to the Sea"\footnote{Used by both anti-Muslim and anti-Semitic groups to express the desire to eliminate Jews/Muslims completely from the Jordan River to the Mediterranean Sea} and "Palestinian People do not Exist."

We used the CrowdTangle API to identify and collect Facebook group posts containing these keywords. The search was conducted in October and November 2023, after Hamas launched rockets into Israel on October 7, 2023. Three graduate students reviewed the groups and posts to determine if the posts expressed hate and if the groups focused on Middle Eastern or Israel-Palestine issues. They identified 128 groups potentially engaged in hate speech. One author reviewed and annotated the results, labeling these groups as "anti-Palestine/Muslim", "anti-Israel/Semitism", "anti-both", or "other". We collected historical posts from these groups. For this study, the posts are from July 2023 to June 2024, covering 32 groups. These groups produced 550.122 posts during the twelve months.

\subsection{Sentiment Analysis}
After removing stop words from each Facebook post in our dataset, we conducted sentiment analysis using a pre-trained model from Hugging Face, finiteautomata/bertweet -base-sentiment-analysis \citep{perez2021pysentimiento}. This model uses BERTweet as the base model and was trained on the SemEval 2017 corpus. Our sentiment analysis resulted in each post being tagged with either positive, neutral, or negative sentiment.

\subsection{Topic Modeling}

We modeled the Facebook corpus topics using BERTopic \cite{grootendorst2022bertopic} models. The 160,456 English posts are tokenized into sentences for training the BERTopic model. We initialized the semantic module with pretrained sentence BERT and update the class-based TF-IDF on our sentence-level Facebook corpus \cite{grootendorst2022bertopic}. Topic clusters were identified using HDBSCAN \cite{campello2013density} with a minimum cluster size of 1,000 sentences. This process yielded 263 fine-grained topic clusters, which are aggregated into six major clusters based on their semantic centroids and top keywords, as detailed below.

\begin{itemize}
\item 1) \textbf{Social Media Activism}: Includes "call for action" phrases and slogans, such as "ImWithIsrael."
\item 2) \textbf{Religious}: Contains references or metaphors from religious texts like the Quran, Bible, and Tanakh.
\item 3) \textbf{Politics and Protests}: Involves mentions of Western political figures and information about offline protests.
\item 4) \textbf{Military}: Covers reports and discussions on military actions in Gaza, such as bombings or ground assaults.
\item 5) \textbf{Ethnicity-based Hate and War-crime Accusations}: Includes explicit hate towards Muslim or Jewish groups and accusations of war crimes in the conflict.
\item 6) \textbf{Irrelevant Topics}: Contains texts unrelated to the Israel-Palestine conflict, such as social media greetings and advertisements.
\end{itemize}

We applied the trained BERTopic model to the 160,456 posts, creating a topic distribution matrix where each post is represented by a 5-dimensional topic vector (excluding the irrelevant topic). 


\subsection{Major event selection}
We selected key events for our analysis based on the highlighted events from the Wikipedia's page "Timeline of the Israel–Hamas war" \footnote{\url{https://en.wikipedia.org/wiki/Timeline_of_the_Israel-Hamas_war}}. These major events include the initial attack, the invasion of the Gaza Strip, and the first ceasefire. There are two other major events on that list: the Yemen airstrikes and the Rafah offensive. In our exploration, we noticed that the first three major events often coincided with significant changes in the trends of engagement, sentiment, and/or topic, and hence we continue using these three events in our analysis. However, since we rarely observed major changes in our time series regarding the Yemen airstrikes, we omitted this event from our current analysis to increase clarity. Furthermore, we skipped the Rafah offensive that started on May 7, 2024, because we missed 10 days of data following this event, hindering our ability to determine trends around that period. Instead, we added two major events that often coincided with significant changes in our time series: the major attack on Rafah \citep{jpost2024} and the withdrawal of Israeli troops from Khan Yunis \citep{cnn2024}. The list of these major events with the corresponding dates is available in Table \ref{majorEvents}.

\vspace{-0.5cm}
\begin{table}[ht]
\caption{Key events related to the recent Israel-Hamas we used in our analysis.}\label{tab1}
\centering
\renewcommand{\arraystretch}{1.25} 
\begin{tabular}{p{0.08\textwidth} p{0.25\textwidth} p{0.6\textwidth}} 
\toprule
\textbf{No.} & \textbf{Date} & \textbf{Event}\\
\midrule
1. & October 7, 2023 & Hamas attacked Israel.\\
2. & October 28, 2023 & Israel started the invasion of the Gaza Strip.\\
3. & November 24, 2023 & The seven-day ceasefire formally started.\\
4. & December 15, 2023 & Israel initiated the first major assault on Rafah (Gaza). \\
5. & April 7, 2024 & Israeli troops withdrew from Khan Yunis (Gaza).\\
\bottomrule
\label{majorEvents}
\end{tabular}
\end{table}
\vspace{-35pt}

\section{Analysis}

\subsection{Trends in Social Media Engagement.}

\begin{figure} [ht]
\includegraphics[width=\textwidth]{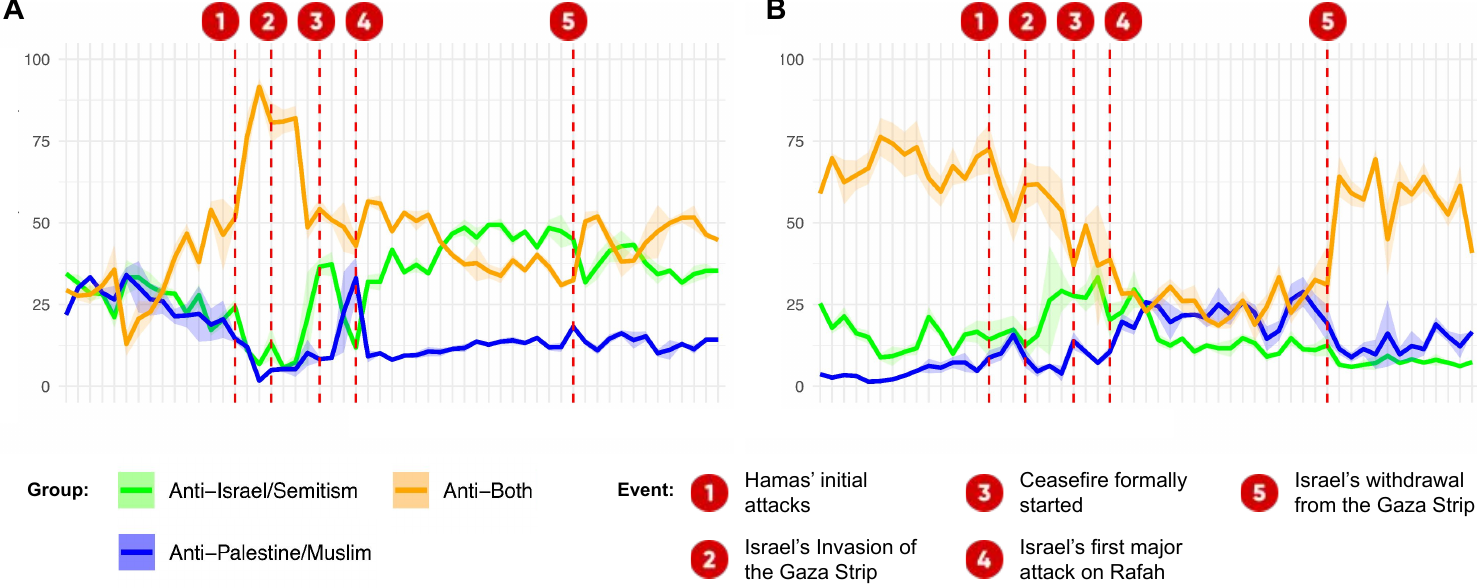}
\caption{The engagement of Facebook groups from July 1st, 2023 to June 30, 2024: (A) the proportions of posts over time, and (B) the proportions of interactions per post over time. The data are aggregated weekly. Solid lines are medians, while shaded areas are 95\% CI. Statistics are normalized to percentages for anti-Israel/Semitism, anti-Palestine/Muslim, anti-both, and anti-others. The anti-others category is omitted for clarity.} \label{engagement}
\end{figure}

\paragraph{\textbf{
Initial Engagement Dynamics and Conflict Onset.}}
From July 1, 2023, to June 30, 2024, the engagement dynamics in Facebook groups showed notable fluctuations, as depicted in Figure~\ref{engagement}. 
Figure~\ref{engagement}A highlights the distribution of post proportions across different group categories over time. While the proportions of posts across the three group categories were similar around mid-2023, a few weeks before Hamas's attack on October 7th, the proportion of posts from the anti-both groups started to increase steeply. At the start of the latest war, the proportion of posts from the anti-both groups was the highest. However, the proportion of posts from the anti-both groups decreased significantly with the onset of the conflict, marked by Hamas's initial attack and Israel's subsequent invasion of Gaza. 

This gradual decrease in the proportion of posts from the anti-both group was accompanied by a significant rise in the proportion of posts from anti-Israel/anti-Semitism groups. Albeit not as sharply as that from anti-Israel/anti-Semitism groups, the proportion of posts from anti-Palestine/anti-Muslim groups also gradually rose. It is possible that, with the onset of the war, individuals focused their attention more on one side than on both sides. Future research should explore these phenomena further to test this hypothesis.

Some significant changes were also observed in the percentage of interactions received per post (Figure \ref{engagement}B). 
A few weeks before the war, posts from the anti-both group category received the highest interactions, outpacing those from the anti-Israel/anti-Semitism and anti-Palestine/Muslim categories. However, after the onset of the war, the proportion of posts from the anti-both groups decreased sharply, while those from the anti-Israel/anti-Semitism and anti-Palestine/Muslim groups generally increased, despite some fluctuations.

Another interesting change during the war was the reversal of interaction patterns: before the war, anti-Israel/Semitism groups had a higher interaction per post proportion compared to anti-Palestine/Muslim groups. This pattern reversed after the onset of the war, particularly after the Rafah attack, with posts from anti-Palestine/Muslim groups receiving more interactions than those from anti-Israel/Semitism groups. This pattern in interaction proportions during the war contrasts with the pattern in post proportions: the proportion of posts from anti-Israel/Semitism groups was higher than that of anti-Palestine/Muslim groups. This may indicate that during the war, users in anti-Israel/Semitism groups were more focused on creating posts than those in anti-Palestine/ Muslim groups, but the opposite was true in terms of generating interactions.

\paragraph{\textbf{Reversals in Engagement Trends Around Israel's Troop Withdrawal.} }
After the start of the war, the proportion of posts from anti-Israel/Semitic groups was the highest. However, around the time of Israel's withdrawal from Khan Yunis, the pattern reversed, with a slight decrease in the proportion of posts from anti-Israel/Semitic groups and a slight increase in the proportion of posts from the anti-both groups.

Meanwhile, during the war, the interactions per post among the anti-both groups and anti-Palestine/Muslim groups became nearly equal, at around 25\%. The proportion of interactions per post from the anti-Israel/Semitic groups was slightly lower than those of the anti-both and anti-Palestine/Muslim groups. This distribution quickly changed around Israel's withdrawal from Khan Yunis. At this point, the interactions per post among the anti-both groups rose significantly, while the proportions from the anti-Palestine/Muslim and anti-Israel/anti-Semitic groups decreased. These changes during Israel's troop withdrawal reflect patterns opposite to those observed at the onset of the war.


\paragraph{\textbf{The Post Rates in Different Groups During the Ceasefire.}}
The ceasefire that began on November 24 was associated with a decline in the percentage of posts from anti-both groups. 
At the same time, the growth rate of posts from anti-Israel/Semitism groups nearly stopped and significantly reduced one week later. The ceasefire was also followed by an increase in posts from anti-Palestine/anti-Muslim groups a week later. 
This pattern may suggest that the ceasefire, which temporarily halted Israel's invasion, reduced engagement proportion in anti-both groups and prevented a surge in negative views toward Israel/Jews, leading to an increase in negative view proportion toward Palestine/Muslims.

\paragraph{\textbf{Variation in Group Posts During Major Military Actions.}}
Israel's major attack on Rafah on December 15 was associated with a notable shift: the proportion of posts from anti-both groups increased by approximately 12.5\%. 
Post proportion from anti-Israel/Semitism groups also saw a rise of almost 20\%, while those from anti-Palestine/ Muslim groups decreased by ~25\%. 
Previously, around the time of Israel's first attack on Gaza on October 28, we observed a marginal increase in anti-both group posts as well. However, this slight increase was accompanied by a marginal decrease in posts from anti-Israel/anti-Semitism groups.
We conjecture that the first attack on Gaza was perceived as more justified due to its nature as a response to a Hamas attack, resulting in a slight decrease in the proportion of posts in anti-Israel/anti-Semitism groups compared to the significant increase observed when Israel launched its first major attack on Rafah.

\subsection{Sentiment Analysis Across Different Facebook Groups}
\begin{figure} [ht]
\includegraphics[width=\textwidth]{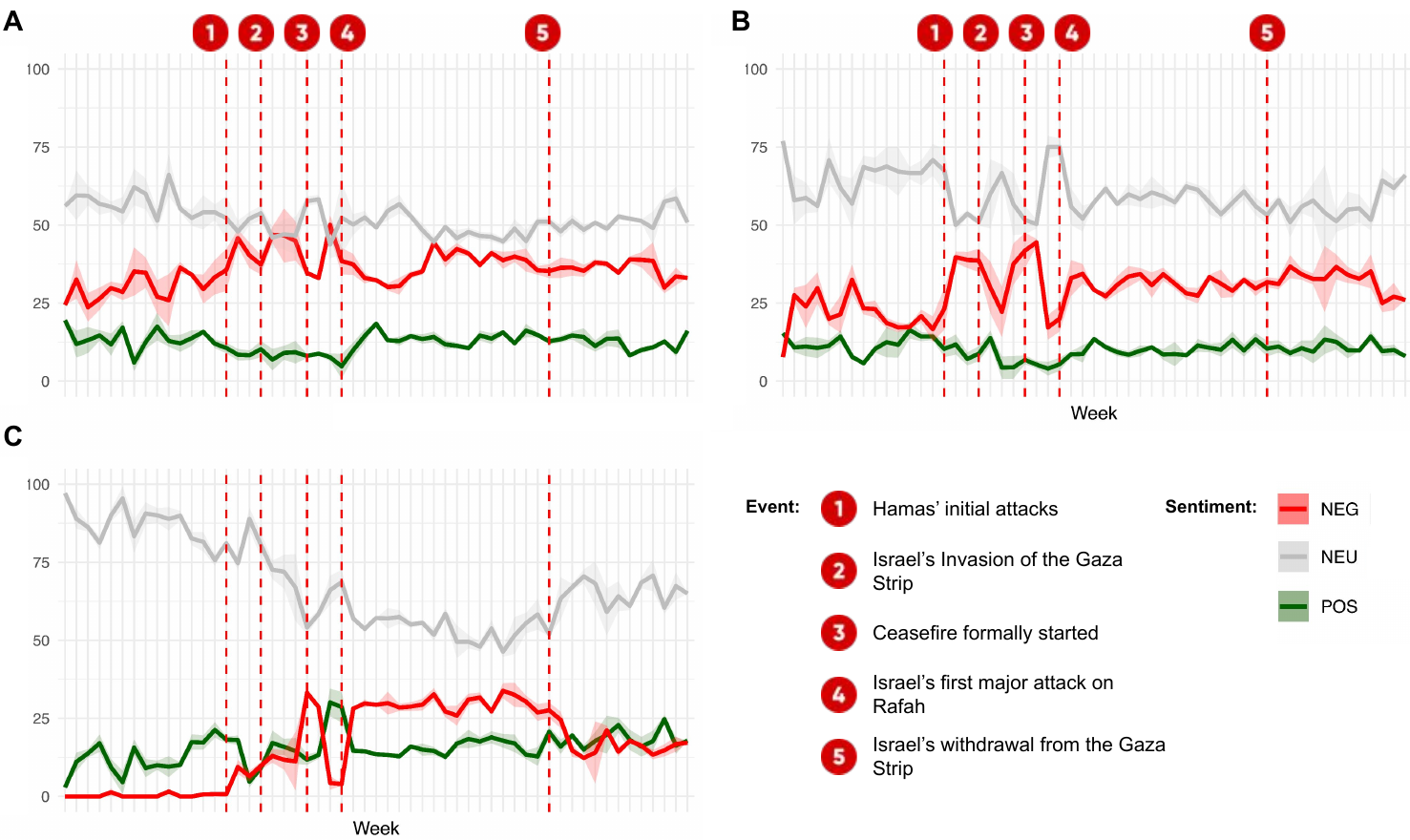}
\caption{The percentages of posts with their corresponding sentiment on the Facebook groups over time (from July 1st, 2023, to June 30, 2024): (A) from anti-Israel/Semitic groups, (B) from anti-Palestine/Muslim groups, and (C) from anti-both groups. The data are aggregated weekly. The solid lines represent the medians, while the shaded areas represent the 95\% CI.} \label{sentiment}
\end{figure}

\paragraph{\textbf{Anti-Israel/Semitism Groups: Predominantly Negative Sentiments.}}
Figure~\ref{sentiment}A illustrates the sentiments over time within anti-Israel/Semitism groups. 
These groups generally exhibited less neutral and more negative sentiments compared to anti-Palestine/ Muslim groups (Figure~\ref{sentiment}B). 
The onset of the conflict, marked by Hamas's attack and Israel's responses, coincided with an increase in negative sentiment and a decrease in neutral sentiments within both group categories.

\paragraph{\textbf{Anti-Both Groups: Drastic Shifts in Sentiment.}}
Figure~\ref{sentiment}C displays the sentiment trends in anti-both groups, which experienced more dramatic changes compared to the other two categories. 
These groups were more neutral before the conflict than other groups, but the neutrality significantly decreased during the war, accompanied by a sharp increase in negativity. 
Following the withdrawal of Israeli forces from Khan Yunis, neutrality increased while negativity decreased.

\paragraph{\textbf{Major Events and Sentiment Trends.}}
Significant events coincided with these trends moving in opposite directions.
For example, Israel's invasion of Gaza on October 28 corresponded with a rise in negative sentiments in the anti-Israel/Semitism groups and a decrease in negative sentiments in the anti-Palestine/Muslim groups. 
Differently, the ceasefire was associated with a slight decrease in negative sentiments within anti-Israel/Semitism groups and a slight increasing trend in negative sentiments within anti-Palestine/Muslim groups.

Regarding the anti-both groups, the proportion of positive sentiment was higher than that of negative sentiment before the war. 
However, this pattern reversed during the war. 
By the end of our study period, the proportion of positive sentiment had marginally surpassed negativity once again, albeit by a very small difference.

\subsection{Topic Trends Across Different Facebook Groups}
\label{sec:result_topic}
\begin{figure} [ht]
\includegraphics[width=\textwidth]{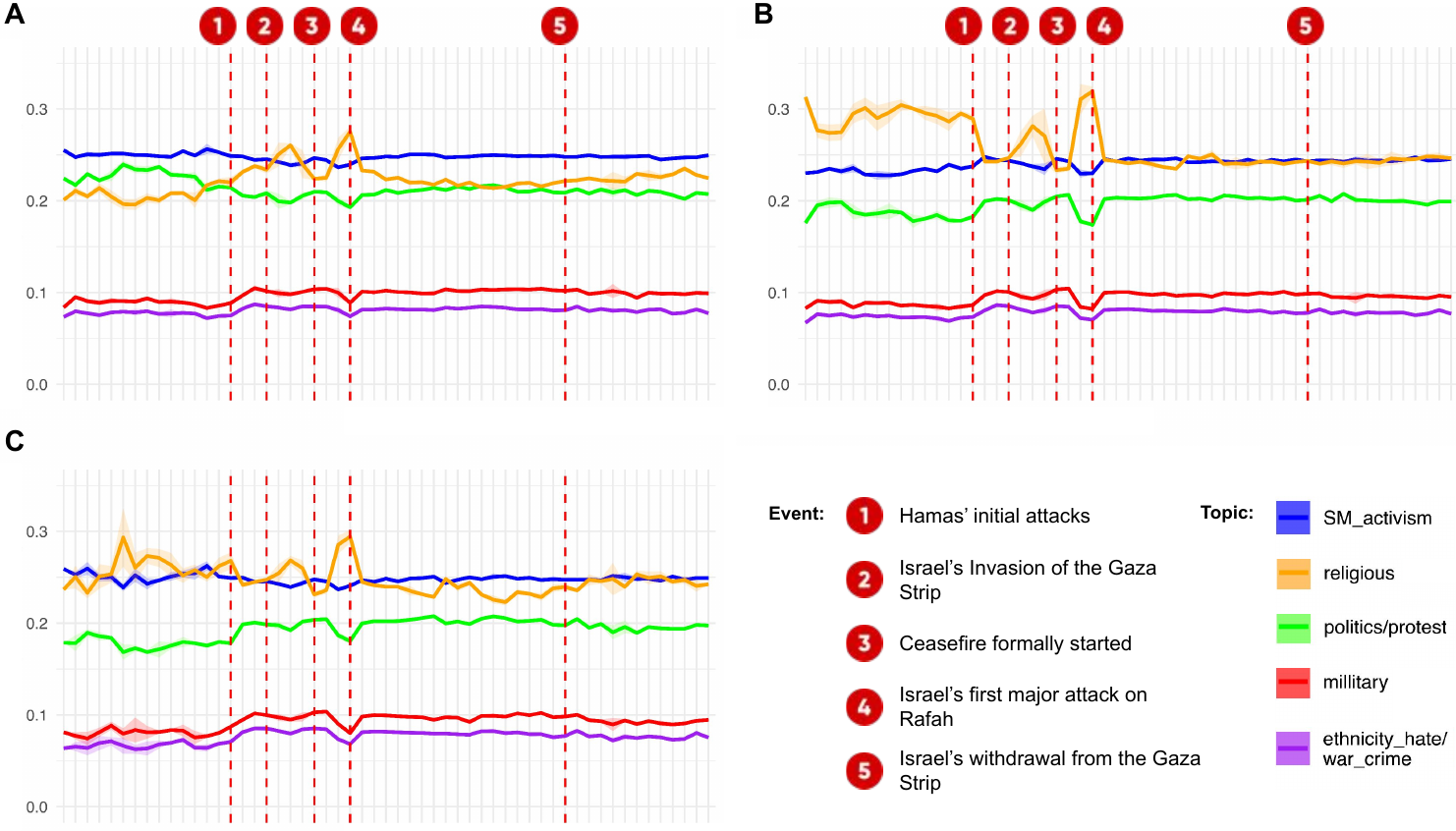}
\caption{The distribution of topics on the Facebook groups over time (from July 1st, 2023, to June 30, 2024): (A) in anti-Israel/Semitic groups, (B) in anti-Palestine/Muslim groups, and (C) in anti-both groups. The data are aggregated weekly. The solid lines represent the medians, while the shaded areas (which are not too apparent due to the small daily variability) represent the 95\% CI).} \label{topic}
\end{figure}

Figure~\ref{topic} breaks down the trend of daily topic probabilities across various Facebook group categories and overall from July 1, 2023, to June 30, 2024.

\paragraph{\textbf{Anti-Israel/Semitism Groups: Social Media Activism and Religious Topics.}}
In the anti-Israel/Semitism groups (Figure \ref{topic}A), in general, the most prominent topic discussed was social media activism, followed by religious topics.
Before the conflict began, these groups focused more on political and protest topics than religious ones. 
However, this trend shifted approximately two weeks before the Hamas attack on October 7, with religious topics becoming more prevalent than political discussions.

\paragraph{\textbf{Anti-Palestine/Muslim Groups: Predominance of Religious Discussions.}}
In contrast, the anti-Palestine/Muslim groups (Figure \ref{topic}B) predominantly discussed religious topics, especially before the first major attack on Rafah. 
Social media activism was the next most discussed topic. 
After the first major attack on Rafah, there was a noticeable decline in religious discussions within these groups, while social media activism and political/protest topics gained a little more prominence.

\paragraph{\textbf{Anti-Both Groups: Shifts in Discussion Topics.}}
Among the anti-both groups (Figure \ref{topic}C), religious and social media activism topics were discussed the most and at similar rates overall. 
However, religious topics were slightly more prevalent before the first major attack on Rafah. 
Between the attack on Rafah and the withdrawal of Israel's troops from Gaza, social media activism topics were discussed more than religious ones. After the Israel's troops withdrawal, the probabilities of the two topics became very similar. Similar to anti-Palestine/Muslim groups, political/protest topics ranked third as the most prominent topics discussed in the anti-both groups and began to be discussed more frequently when the war started.

\subsection{Qualitative Answers for the RQs}

The analysis conducted in this study provides several insights into the research questions posed. 
Regarding \textbf{RQ1} (How does engagement with hate speech vary as the Israel-Palestine conflict develops?), the data indicates that engagement levels fluctuated significantly in response to key events. 
Notably, the onset of the conflict was associated with an increase in single-sided posts (either anti-Israel/anti-Semitism or anti-Palestine/Muslim), suggesting a polarization in user engagement. 
For \textbf{RQ2} (How do sentiments within anti-Muslim and anti-Jew groups change during the conflict?), the data reveals that both groups exhibited a marked increase in negative sentiments coinciding with major conflict events, such as Hamas’s initial attack and Israel's invasion of Gaza.
However, the ceasefire period saw a complex shift, with a decrease in negative sentiments within anti-Israel/anti-Semitism groups and an increase in anti-Palestine/Muslim groups. 
Lastly, for \textbf{RQ3} (How do the topics of social media discussions change within these groups over time?), the analysis highlights that the dominant topics within these groups shifted notably during the conflict. 
Anti-Israel/Semitism groups transitioned from discussions on political/protest topics to religious ones, although this change started before the war.
Anti-Palestine/Muslim groups initially concentrated on religious topics, which decreased after the first major attack on Rafah, giving way to more social media activism and political/protest discussions. 
These findings shows the dynamics of online hate speech targeting Muslims and Jews, and the relationships between geopolitical events and the narratives within groups.

\section{Discussion and Conclusion}

In this study, we examined the dynamics of Facebook groups from July 1, 2023, to June 30, 2024, focusing on engagement, sentiments, and topics discussed. We classified these groups into four categories: anti-Israel/Semitism, anti-Palestine/Muslim, anti-both, and other. We then analyzed how trends in the first three categories corresponded with five major events of the Israel-Hamas war.

Our findings support the hypothesis that Facebook group dynamics coincided with major events. The onset of the conflict, marked by Hamas's attack and subsequent Israeli actions, coincided with a decrease in content creation proportion by anti-both users and an increase among anti-Israel/Semitism and anti-Palestine/Muslim users. This pattern reversed after the Israeli troop withdrawal from Gaza, with activity proportion rising in the anti-both groups and falling in the other two categories. During the war's start, negative content percentage increased, and neutral content percentage decreased in the all group categories. Anti-Palestine/Muslim groups shifted from religious to social media activism and political/protest discourse, while anti-Israel/Semitism groups moved from political/protest to religious topics, although this shift began before the war.


These findings have several implications. The coinciding changes in engagement, sentiment, and topic trends suggest a potential correlation between online and offline expressions of hate. This provides a basis for a more refined time series analysis in the future.
The Facebook groups exhibited more negativity and less neutrality during the war, indicating a shift in discourse patterns.
It suggests that the social media platforms may implement effective strategies to combat negativity without impeding constructive criticism, while it poses challenges for community moderators to carefully differentiate between critical feedback and hate speech.

\subsection{Limitation and Future Works}

This study has several limitations, which also point to directions for future research. The dataset covers only one year, providing a limited view of the conflict's complex and evolving nature. 
The qualitative approach, focusing on event coincidences, restricts our understanding of causal relationships. 
Future research should combine quantitative and qualitative methods to explore these relationships more thoroughly. 
Additionally, while this analysis grouped anti-Israel with anti-Semitism and anti-Palestine with anti-Muslim views for simplicity, future studies should aim to distinguish between these concepts more precisely to better understand the dynamics of hostility toward these groups. 
Lastly, regarding the methodological consideration, the graduate students and the reviewer who examined the groups and posts to assess whether they expressed hate speech were not from either country involved in the conflict. While this likely helped reduce potential biases in their review, we recognize that personal biases, if any, can never be entirely eliminated, which remains a limitation of the study.

\begin{credits}

\subsubsection{\discintname}
The authors declare no conflicts of interest. While it is acknowledged that personal backgrounds may have some potential for influence on the research, the analysis was conducted with a focus on impartiality.

\subsubsection{Acknowledgement}
The authors would like to acknowledge support from AFOSR, ONR, Minerva, NSF \#2318461, and Pitt Cyber Institute's PCAG awards. The research was partly supported by Pitt's CRC resources. Any opinions, findings, and conclusions or recommendations expressed in this material do not necessarily reflect the views of the funding sources.
\end{credits}

%
%
%
%

\renewcommand{\bibsection}{%
  \section*{References}%
  \footnotesize 
}
\bibliographystyle{splncs04}
\bibliography{references}

\end{document}